%% file: nlevel.tex
\documentclass[12pt]{article}
\usepackage{fullpage}

\newcommand{\Variance}{\sigma^2}

\newcommand{\expansion}{\mathrm{expansion}}
\newcommand{\algname}{KaSPar}
\usepackage{subfigure}
\usepackage{lscape}

\usepackage{url}
\usepackage{color}
\usepackage{epic}
\usepackage{eepic}
\usepackage{amsmath}
\usepackage{graphicx}
\input{makros2.tex}

\def\MdR{\ensuremath{\mathbf{R}}}

\newcommand{\postponed}[1]{}
\renewcommand{\frage}[1]{}

\newcommand{\partition}{\mathcal{P}}
\newcommand{\gain}[2]{g_{#1}(#2)}

\title{$n$-Level Graph Partitioning\thanks{Partially supported by DFG grant SA 933/3-2.}}
\author{Vitaly Osipov\thanks{Universit\"at Karlsruhe (TH), Germany, {\tt osipov@ira.uka.de}} \and Peter Sanders\thanks{Universit\"at Karlsruhe (TH), Germany, {\tt sanders@ira.uka.de}}}
\date{}

\begin{document}
\maketitle
\pagestyle{plain}

\begin{abstract}
 We present a multi-level graph partitioning algorithm based on the extreme   idea to contract only a single edge on each level of the hierarchy.  This   obviates the need for a matching algorithm and promises very good partitioning   quality since there are very few changes between two levels. Using an   efficient data structure and new flexible ways to break local search   improvements early, we obtain an algorithm that scales to large inputs and       produces the best known partitioning results for many inputs. For example, in Walshaw's well known benchmark tables we achieve 155 improvements dominating the entries for large graphs.
\end{abstract}

\section{Introduction}
\label{s:intro}
Many important applications of computer science involve processing large graphs,
e.g., stemming from finite element methods, digital circuit design, route
planning, social networks, etc. Very often these graphs need to be partitioned
or clustered such that there are few edges between the blocks (pieces).

\begin{figure}[t]
\begin{center}
\includegraphics[width=0.8\textwidth]{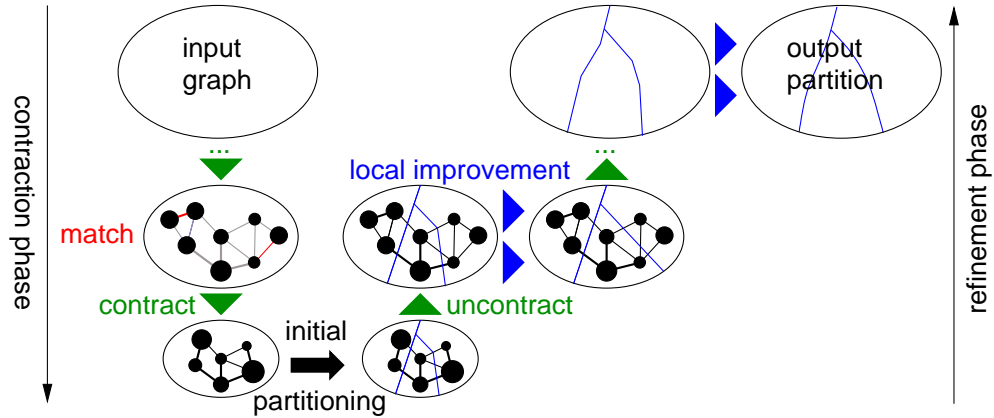}
\end{center}
\vspace{-7mm}
\caption{Multilevel graph partitioning.}
\label{fig:mgp}
\end{figure}

A successful heuristic for partitioning large graphs is the \emph{multilevel graph partitioning}
approach (MGP) depicted in Figure~\ref{fig:mgp} where the graph is recursively
\emph{contracted} to a smaller graph with the same basic
structure. After applying an \emph{initial partitioning} algorithm to this small
graph, the contraction is undone\postponed{it is uncontracted again} and, at each level, a \emph{local refinement}
method improves the partition induced by the coarser level.
Section~\ref{s:preliminaries} explains the method in more detail.  Most systems
instantiate MGP in a very similar way: Maximal matchings are contracted between
two levels that try to include as many heavy edges as possible. 
Local refinement uses a linear time variant of local search. MGP has two crucial advantages over most other
approaches to graph partitioning: We get near linear execution time since the
graph shrinks geometrically and we get good partitioning quality since a good
solution on some level yields a good initial solution on the next finer level,
i.e., local search needs little work to further improve the solution.

Our central idea is to get even better partitions by making subsequent levels as similar as possible -- we (un)contract only a \textit{single} edge between two levels.  We call this $n$-GP since we have (almost) $n$ levels of hierarchy. More details are described in Section~\ref{s:nMGP}.  $n$-GP has the additional advantage that there is no longer a need for an algorithm finding heavy matchings. This is remarkable insofar as a considerable amount of work on approximate maximum weight matching was motivated by the MGP application \cite{Preis99,DH03c,PS04,MauSan07}. Still, at first glance, $n$-GP seems to have substantial disadvantages also. Firstly, storing each level explicitly would lead to quadratic space consumption. We avoid this by using a dynamic graph data structure with little space overhead.  Secondly, choosing maximal matchings instead of just a single edge for contraction has the side effect that the graph is contracted everywhere, leading to a more uniform distribution of node weights.  We solve this problem by explicitly factoring node weights into the \emph{edge rating} function prioritizing the edges to be contracted. Already in \cite{DAHoltgrewe,HSS10} edge ratings have proven to lead to better results for graph partitioning. Perhaps the most serious problem is that the most common approach to local search is to let it run for a number of steps proportional to the current number of nodes. In the context of $n$-GP this could lead to a quadratic overall number of local search steps. Therefore, we develop a new, more adaptive stopping criteria for the local search that drastically accelerates $n$-GP without significantly reducing partitioning quality.

We have implemented $n$-GP in the graph partitioner \algname\ (Karlsruhe
Sequential Partitioner). Experiments reported in Section~\ref{s:experiments}
indicate that \algname\ scales well to large networks, computes the best known
partitions for many instances of a ``standard benchmark'' and needs time
comparable to system that previously computed the best results for large
networks.  Section~\ref{s:conclusion} summarizes the results and discusses
future directions.

\subsection*{More Related Work}
There has been a huge amount of research on graph partitioning so that we refer
to introductory and overview papers such as
\cite{fjallstrom1998agp,KarKum98b,SchKarKum00,Walshaw07} for more
material. Well-known software packages based on MGP are Chaco
\cite{Chaco}, DiBaP \cite{mms-a-09},  Jostle \cite{jostleS,Walshaw07},
Metis~\cite{metisS,metisM},
Party \cite{partyS,partyM}, and Scotch~\cite{p-scotc-07,p-scotc-08}.

\algname\ was developed partly in parallel with KaPPa (Karlsruhe Parallel
Partitioner) \cite{HSS10}.  KaPPa is a ``classical'' matching based MGP
algorithm designed for scalable parallel execution and its local search only
considers independent pairs of blocks at a time. Still, for $k=2$, its
interesting to compare \algname\ and KaPPa since KaPPa achieves the previously
best partitioning results for many large graphs, since both systems use a
similar edge ratings, and since running times for a two processor
parallel code and a sequential code could be expected to be roughly comparable.

There is a long tradition of $n$-level algorithms in geometric data structures
based on randomized incremental construction (e.g, \cite{GKS92,BHSS10short}).  Our
motivation for studying $n$-level are \emph{contraction hierarchies}
\cite{GSSD08}, a preprocessing technique for route planning that is at the same
time simpler and an order of magnitude more efficient than previous techniques
using a small number of levels.

\section{Preliminaries}\label{s:preliminaries}

Consider an undirected graph $G=(V,E,c,\omega)$ 
with edge weights $\omega: E \to \MdR_{>0}$\postponed{todo find a pdflatex replacement for mathbb font}, node weights
$c: V \to \MdR_{\geq 0}$, $n = |V|$, and $m = |E|$.
We extend $c$ and $\omega$ to sets, i.e.,
$c(V')\Is \sum_{v\in V'}c(v)$ and $\omega(E')\Is \sum_{e\in E'}\omega(e)$.
$\Gamma(v)\Is \setGilt{u}{\set{v,u}\in E}$ denotes the neighbors of $v$.

We are looking for \emph{blocks} of nodes $V_1$,\ldots,$V_k$ that partition $V$,
i.e., $V_1\cup\cdots\cup V_k=V$ and $V_i\cap V_j=\emptyset$ for $i\neq j$. The
\emph{balancing constraint} demands that $\forall i\in 1..k\gilt c(V_i)\leq
L_{\max}\Is (1+\epsilon)c(V)/k+\max_{v\in V} c(v)$ for some parameter
$\epsilon$.  The last term in this equation arises because each
  node is atomic and therefore a deviation of the heaviest node has to be
  allowed.  The objective is to minimize the total \emph{cut}
$\sum_{i<j}w(E_{ij})$ where $E_{ij}\Is\setGilt{\set{u,v}\in E}{u\in V_i,v\in
  V_j}$.  By default, our initial inputs will have unit edge and node weights.
However, even those will be translated into weighted problems in the course of
the algorithm.

\emph{Contracting} an edge $\set{u,v}$ means replacing the nodes $u$ and $v$ by
a new node $x$ connected to the former neighbors of $u$ and $v$. We set
$c(x)=c(u)+c(v)$. If replacing edges of the form $\set{u,w},\set{v,w}$ would
generate two parallel edges $\set{x,w}$, we insert a single edge with
$\omega(\set{x,w})=\omega(\set{u,w})+\omega(\set{v,w})$.  \emph{Uncontracting}
an edge $e$ undoes its contraction. Partitions computed for the contracted graph
are extrapolated to the uncontracted graph in the obvious way, i.e., $u$ and $v$
are put into the same block as $x$.

\emph{Local Search} is done by moving single nodes between blocks. The gain 
$\gain{B}{v}$ of moving node $v$ to block $B$ is decrease in total cut size caused by this move.
For example, if $v$ has 5 incident edges of unit weight, 2 of which are inside $v$'s block
and 3 of which lead to block $b$ then $\gain{B}{v}=3-2=1$

\section{$n$-Level Graph Partitioning}\label{s:nMGP}

\begin{figure}
\begin{code}
  \Function $n$-GP$(G, k, \epsilon)$\+\\
  \If $G$ is small \Then \Return initialPartition$(G,k, \epsilon)$\\
  pick the edge $e=\set{u,v}$ with the highest rating\\
  contract $e$;\quad
  $\partition\Is n\Id{-GP}(G, k, \epsilon)$;\quad
  uncontract $e$\\
  activate$(u)$;\quad activate$(v)$;\quad localSearch$()$\\
  \Return $\partition$
\end{code}
\vspace{-7mm}
\caption{$n$-GP.\label{fig:nGP}}
\end{figure}

\postponed{rewritten:}Figure~\ref{fig:nGP} gives a high-level recursive summary of $n$-GP.  The base case is some other partitioner used when the graph is sufficiently small. In \algname, contraction is stopped when either only $20k$ nodes remain, no further nodes are eligible for contraction, or there are less edges than nodes left. The latter happens when the graph consists of many independent components. As observed in \cite{HSS10} Scotch~\cite{p-scotc-07} produces better initial partitions than metis, and therefore we also use it in \algname\ .

The edges to be contracted are chosen according to an edge rating function. \algname\ adopts the rating function
$$\expansion^*(\set{u,v})\Is \frac{\omega(\set{u,v})}{c(u)c(v)}$$
which fared best in \cite{HSS10}. As a further measure to avoid
unbalanced inputs to the initial partitioner, \algname\ never allows
a node $v$ to participate in a contraction if the weight of $v$ exceeds 
$1.5n/(20k)$.
Selecting contracted edges can be implemented efficiently by keeping the contractable \emph{nodes} in a priority queue sorted by the rating of their 
most highly rated incident edge. 

In order to make contraction and uncontraction efficient, we use a ``semidynamic'' graph data structure: When contracting an edge $\set{u,v}$, we mark both $u$ and $v$ as deleted, introduce a new node $w$, and redirect the edges incident to $u$ and $v$ to $w$. 
The advantage of this implementation is that edges adjacent to a node are still stored in adjacency arrays which are more efficient than linked lists needed for a full fledged dynamic graph data structure. A disadvantages of our approach is a certain space overhead. However, it is relatively easy to show that this space overhead is bounded by a logarithmic factor even if we contract edges in some random fashion (see \cite{DSSS04}). In Section~\ref{s:experiments} we will demonstrate experimentally that the overhead is actually often a small constant factor. Indeed, this is not very surprising since the edge rating function is not random, but designed to keep the contracted graph sparse.  Overall, with respect to asymptotic memory overhead, $n$-GP is no worse than methods with a logarithmic number of levels.

\subsection{Local Search Strategy}

Our local search strategy is similar to the FM-algorithm  \cite{fiduccia1982lth} that is also
used in many other MGP systems. We now outline our variant and then discuss differences. 
\par Originally, all nodes are unmarked. Only unmarked nodes are allowed to be activated or moved from one block to another. 
Activating a node $v \in B'$  means that for  blocks 
$\setGilt{ B\neq B'}{\exists \set{v,u}\in E \wedge u\in B }$
we compute the gain
$$\gain{B}{v}=\sum\setGilt{\omega(\set{v,u})}{\set{v,u}\in E, v\in B}-\sum\setGilt{\omega(\set{v,u})}{\set{v,u}\in E, v\in B'}$$ 
of moving $v$ to block $B$. 
Node $v$ is then inserted into the priority queue $P_B$
using $\gain{B}{v}$ as the priority.  We call a queue $P_B$ eligible if the
highest gain node in $P_b$ can be moved to block $B$ without violating the
balance constraint for block $B$.  Local search repeatedly looks for the highest
gain node $v$ in any eligible priority queue $P_B$ and moves $v$ to block
$B$. When this happens, node $v$ becomes nonactive and marked, the unmarked neighbors of $v$ get activated and the
gains of the active neighbors are updated. 
The local search is stopped if
either no eligible nonempty queues remain, or one of the stopping criteria
described below applies.  After the local search stops, it is rolled back to the
lowest cut state reached during the search (which is the starting state if no
improvement was achieved).  Subsequently all previously marked nodes are
unmarked.  The local search is repeated until no improvement is achieved. 

The main difference to the usual
FM-algorithm is that our routine performs a highly localized search starting
just at the uncontracted edge. Indeed, our local search does nothing if none of
the uncontracted nodes is a \emph{border node}, i.e., has a neighbor in another
block.  Other FM-algorithms initialize the search with all border nodes. In
$n$-GP the local search may find an improvement
quickly after moving a small number of nodes.  However, in order to exploit this
case, we need a way to stop the search much earlier than previous
algorithms which limit the number of steps to a constant fraction of the current
number of nodes $|V|$.

\paragraph*{Stopping Using a Random Walk Model.}
\postponed{It is desirable to stop a local search early if it tends to result in worse and worse cuts. On the other hand, a local search that is still likely to give an improvement should be continued further. Therefore, we made an adaptive stopping rule that dependents on ...}%
It makes sense to make a stopping rule more adaptive by making it dependent on
the past history of the search, e.g., on 
the difference between the current cut and the best cut achieved before.

We model the gain values in each step as identically distributed, independent
random variables whose expectation $\mu$ and Variance $\Variance$ is obtained
from the previously observed $p$ steps.  
In Appendix~\ref{app:stop} we show how from these (purely heuristical, i.e., technically unwarranted) assumptions we can derive that it is unlikely that the local search will produce a better cut if 
\begin{equation}p\mu^2>\alpha\Variance+\beta\label{eq:randomwalk}\end{equation}
where $\alpha$ and $\beta$ are tuning parameters and $\mu$ is the average gain since the last improvement. For the variance $\Variance$, we can also use
the variance observed throughout the current local search.
Parameter $\beta$ is a base value that avoids stopping just after a small constant number of steps that happen to have small variance. Currently we set it to $\ln n$.


\section{Trial Trees}\label{s:trialTree}

\frage{rewritten/shortened}It is a standard technique in optimization heuristics to improve results by repeating various parts of the algorithm. We generalize several approaches used in MGP by adapting an idea initially used in a fast randomized min-cut algorithm \cite{KarSte96}: After reducing the number of nodes by a factor $c$, we perform two independent trials using different random seeds for tie breaking during contraction, initial partitioning, and local search. Among these trials the one with the smaller cut is used for continuing upwards. This way, we perform independent trials at many levels of contraction controlled by a single tuning parameter $c$. As long as $c>2$, the total number of contraction steps performed stays $\Oh{n}$.


\section{Experiments}\label{s:experiments}

\paragraph*{Implementation.}
We implemented \algname\ in C++ using gcc-4.3.2. \postponed{todo: 1--2 sentences on data structure and impl. of contraction.}  We use priority queues based on paring heaps \cite{TavoryDK04} available in the policy-based elementary data structures library (pb\_ds) for implementing contraction and refinement procedures.
In the following experimental study we compared \algname\ to Scotch 5.1, kMetis 4.0 and the same version of KaPPa as in \cite{HSS10}. 

\paragraph*{System.}
We performed our experiments on a single core of an Intel Xeon Quad-core Processor featuring 2x4~MB of L2 cache and clocked at 2.667~GHz of a 2 processor Intel Xeon X5355 node with $16$~GB of RAM running Suse Linux Enterprise 10.
\begin{table}[t]
\caption{
  Basic properties of the graphs from our 
  benchmark set. The large instances are split into five groups:
  geometric graphs, FEM graphs, street networks, sparse matrices, and social networks.  Within their groups, the graphs are sorted by size.  }
\begin{center}
\begin{tabular}{|l|r|r|}
\hline
\multicolumn{3}{|l|}{Medium sized instances} \\
\hline
graph & $n$ & $m$ \\
\hline
rgg17	 & $2^{17}$ 	& 1\,457\,506\\
rgg18 	 & $2^{18}$  	& 3\,094\,566\\
Delaunay17	 & $2^{17}$	& 786\,352\\
Delaunay18 	 & $2^{18}$ 	& 1\,572\,792\\
\hline
bcsstk29 	&	13\,992 &	605\,496\\
4elt 		&	15\,606 & 	91\,756\\
fesphere  	& 	16\,386 & 	98\,304\\
cti  		&	16\,840 & 	96\,464\\	
memplus 	&	17\,758 &	108\,384\\
cs4 	 	&       33\,499 & 	87\,716\\
pwt 		&	36\,519 & 	289\,588\\
bcsstk32 	&	44\,609 &	1\,970\,092\\	
body 		& 45\,087 	& 327\,468\\
t60k 		& 60\,005 	& 178\,880\\
wing  		& 62\,032 	& 243\,088\\
finan512 	& 74\,752 	& 522\,240\\	
ferotor		& 99\,617	& 662\,431\\
\hline
bel & 463\,514 & 1\,183\,764\\
nld & 893\,041 & 2\,279\,080\\
\hline
af\_shell9 	& 504\,855 & 17\,084\,020\\\hline
\end{tabular}
\hspace*{\fill}
\begin{tabular}{|l|r|r|}
\hline
\multicolumn{3}{|l|}{Large instances} \\\hline
graph & $n$ & $m$ \\
\hline
rgg20 		& $2^{20}$ 	& 13\,783\,240   \\
Delaunay20 		& $2^{20}$ 	& 12\,582\,744  \\
\hline
fetooth	 	& 78\,136  	& 905\,182\\
598a  		& 110\,971 	& 1\,483\,868\\
ocean 		& 143\,437 	& 819\,186\\
144  		& 144\,649 	& 2\,148\,786\\
wave 		& 156\,317 	& 2\,118\,662  \\
m14b 		& 214\,765 	& 3\,358\,036 \\
auto 		& 448\,695 	& 6\,629\,222 \\
\hline
deu & 4\,378\,446 & 10\,967\,174  \\
eur & 18\,029\,721 & 44\,435\,372 \\
\hline
af\_shell10 & 1\,508\,065 & 51\,164\,260  \\
\hline
\multicolumn{3}{|l|}{Social networks} \\
\hline
coAuthorCiteseer & 227\,320 & 1\,628268 \\
coAutorhDBLP 	& 299\,067 & 1\,955\,352\\       
cnr2000 & 325\,557 & 3\,216\,152 \\ 
citationCiteseer & 434\,102 & 32\,073\,440 \\       
coPaperDBLP & 540\,486& 30\,491\,458\\ 
\hline
\end{tabular}
\end{center}
\label{tab:instances}

\end{table}
\paragraph*{Instances.}
We report results on two suites of instances summarized in
Table~\ref{tab:instances}. \Id{rggX} is a \emph{random geometric graph} with
$2^{X}$ nodes that represent random points in the unit square and edges
connect nodes whose Euclidean distance is below $0.55 \sqrt{ \ln n / n }$.
This threshold was chosen in order to ensure that the graph is almost connected. 
\Id{DelaunayX} is the Delaunay triangulation of $2^{X}$
random points in the unit square.  Graphs \Id{bcsstk29}..\Id{fetooth} and
\Id{ferotor}..\Id{auto} come from Chris Walshaw's benchmark archive
\cite{SWC04plus}.  Graphs \Id{bel}, \Id{nld}, \Id{deu} and \Id{eur} are
undirected versions of the road networks of Belgium, the Netherlands, Germany,
and Western Europe respectively, used in \cite{DSSW09}. Instances
\Id{af\_shell9} and \Id{af\_shell10} come from the Florida Sparse
Matrix Collection \cite{UFsparsematrixcollection}.  \Id{coAuthorsDBLP}, \Id{coPapersDBLP}, 
\Id{citationCiteseer}, \Id{coAuthorsCiteseer} and \Id{cnr2000} are examples of social networks taken from \cite{GSS08}.

For the number of partitions $k$ we choose the values
used in  \cite{SWC04plus}: 2, 4, 8, 16, 32, 64.
Our default value for the allowed imbalance is 3 \% since this is one
of the values used in \cite{SWC04plus} and the default value in Metis.

When not otherwise mentioned, we perform 10 repetitions 
for the small networks and 5 repetitions for the other.
We report the arithmetic average of computed cut size, running time and
the best cut found. When further averaging over multiple instances, we use the geometric
mean in order to give every instance the same influence on the final
figure.

\paragraph*{Configuring the Algorithm.}
We use two sets of parameter settings \emph{fast} and
\emph{strong}. These methods only differ in the
constant factor $\alpha$ in the local search stopping rule, see Equation~(\ref{eq:randomwalk}), in the contraction factor $c$ for the trial tree (Section~\ref{s:trialTree}), and in the number of initial partitioning
attempts $a$ performed at the coarsest level of contraction:
\begin{center}
\begin{tabular}{l|l|c|c}
strategy & $\alpha$ & $c$ & $a$ \\\hline
fast     &  1       &  8  &  $25/\log_2 k$\\
strong   &  4       &  2.5  &  $100/\log_2 k$
\end{tabular}
\end{center}

Note that this are considerably less parameters compared to KaPPa. In particular,
there is no need for selecting a matching algorithm, an edge coloring algorithm, 
or global and local iterations for refinement.

\begin{figure}
\begin{center}
\includegraphics[scale=0.8]{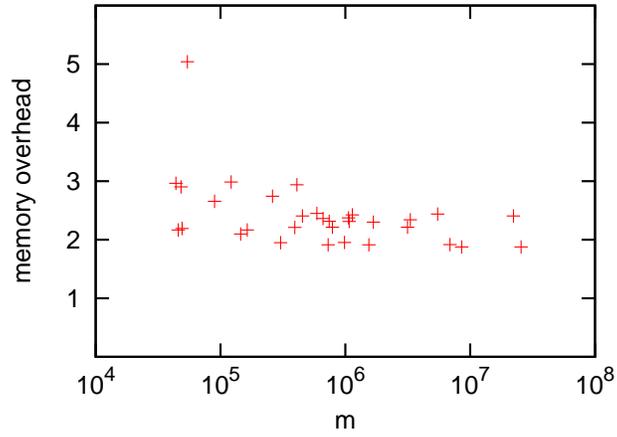}
\end{center}
\vspace*{-7mm}
\caption{Number of edges created during contraction.}
\label{fig:memoryOverhead}
\end{figure}
\begin{figure}
\begin{center}
\includegraphics[scale=0.8]{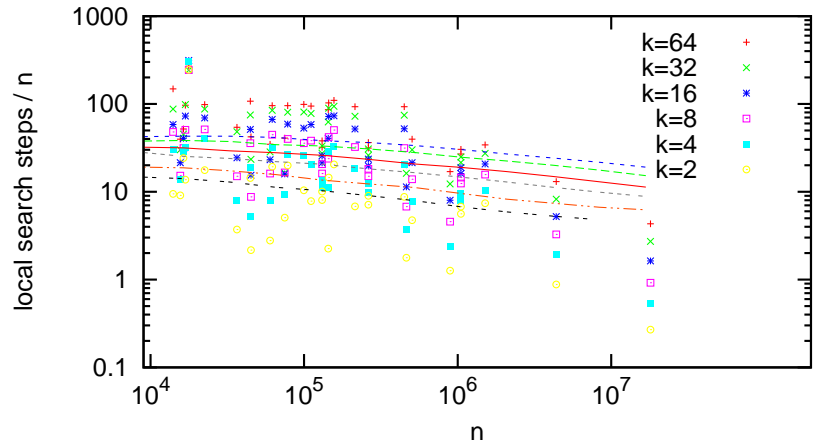}
\end{center}
\vspace*{-7mm}
\caption{Total number of local search steps. The nearly straight lines represent series for the graphs rgg15..rgg24 and Delaunay15..Delaunay24 for different $k$.}
\label{fig:localSearch}
\end{figure}

\paragraph*{Scalability.}
Figure~\ref{fig:memoryOverhead} shows the number of edges touched during
contraction (\algname\ strong, small and large instances).  We see that this scales linearly with the number of input edges
and with a fairly small constant factor between 2 and 3. Interestingly, the
number of local search steps during local improvement (Figure~\ref{fig:localSearch}) \emph{decreases} with
increasing input size. This can be explained by the sublinear number of border
vertices that we have in graphs that have small cuts and by small average search
space sizes for the local search.  Indeed, Figure~\ref{fig:localSearchLength} in
the appendix indicates that the average length of local searches grows only
logarithmically with $n$. All this translates into fairly complicated running time behavior.
Still, Figure~\ref{fig:runTime} in the appendix  
warrants the conclusion that running time scales ``near linearly'' with the input size.\footnote{This may not apply to the social networks which have considerably worse behavior.}
The term in the running time depending on $k$ grows sublinearly with the input size so
that for very large graphs the number of blocks does not matter much.

\paragraph*{Does the Random Walk Model Work?}
We have compared KaSPar fast with a variant where the stopping rule is disabled (i.e., $\alpha=\infty$).  For the small instances this yields about 1 \% better cut sizes at the cost of an order of magnitude larger running time. This is a small improvement both compared to the improvement KaSPar achieves over other systems and compared to just repeating KaSPar fast 10 times (see Table~\ref{tab:geomeans}).

\paragraph*{Do Trial trees help?}
We use the following evaluation: We run \algname\ strong and measure its elapsed time. Then for different values of initial partitionings $a$ we repeat \algname\ strong without trial trees( $c=0$ ), until the sum of the run times of all repetitions exceeds the run time of \algname\ strong. Than for different values $a$ we compare the best edge cut achieved during repeated runs to the one produced by \algname\ strong. Finally, we average the obtained results over 5 repetitions of this procedure. If
we then quality the computed partitions, we usually get almost identical results
(a fraction of a percent difference). However, most of the time trial trees are
a bit better and for \emph{road networks} we get considerable improvements.
For example, for the European network we get an improvement of 10 \% on average over all $k$.

\paragraph*{Comparison with other Systems.}
Table~\ref{tab:geomeans} summarizes the results by computing geometric means
over 10 runs for the small instances and over 5 runs for the large
instances and social networks. We exclude the European road network for
$k=2$ because KaPPa runs out of memory in that case.  Detailed per instance
results can be found in the appendix.  KaPPa strong produces 5.9 \% larger cuts
than KasPar strong for small instances (average value) and 8.1 \% larger cuts for
the large instances.  This comparison might seem a bit unfair because KaPPa is
about five times faster.  However, KaPPa is using $k$ processors in
parallel. Indeed, for $k=2$ KaSPar strong needs only about twice as much time.
Also note that KaPPa strong needs about twice as much time as \algname\ fast
while still producing 6 \% larger cuts despite running in parallel.
The case $k=2$ is also interesting because here KaPPa and KaSPar are most similar --
parallelism does not play a big role (2 processors) and both local search strategies work
only on two blocks at all time. Therefore 6 \% improvement of KaSPar over KaPPa  we can attribute mostly to the larger number of levels.%
\postponed{discuss the effect that the performance ratios KaSPar vs KaPPa strong average behave like 6.3\%, 11.8\%, 10.1\%, 8.5\%, 6.7\%, 5\% for $k=2..64$?}

Scotch and kMetis are much faster than KaSPar but also produce considerably
larger cuts -- e.g., 32~\% larger for large instances (kMetis, average). For the
European road network, the difference in cut size even exceeds a factor of two. Such
gaps usually cannot be breached by just running the faster solver a larger
number of times. For example, for large instances, Scotch is only a factor around
4 faster than KaSPar fast, yet its best cut values obtained from 5 runs are
still 12.7~\% larger than the average values of KaSPar fast.

For social networks all systems have problems. KaSPar lags further behind in
terms of speed but extends its lead with respect to the cut size.  We mostly
attribute the larger run time to the larger cut sizes relative to the number of
nodes which greatly increase the number of local searches necessary. A further
effect may be that the time for a local search step is proportional to the
number of partitions adjacent to the nodes participating in the local search. For ``well
behaved'' graphs this is mostly two, but for social networks
which get denser on the coarser levels this value can get larger.

\begin{table}
\caption{\label{tab:geomeans}Geometric means over all instances.}
\begin{center}
\begin{tabular}{l||r|r|r||r|r|r||r|r|r}
code & \multicolumn{3}{|c||}{small graphs}&\multicolumn{3}{|c||}{large graphs}&\multicolumn{3}{|c}{social networks}\\
              & best   & avg.   & t[s] & best   &  avg.   & t[s] &  best     & avg. & t[s] \\\hline
KaSPar strong & 2\,675 & 2\,729 & 7.37 & 12\,450& 12\,584 & 87.12& -      & - & - \\
KaSPar fast   & 2\,717 & 2\,809 & 1.43 & 12\,655& 12\,842 & 14.43& 93657  &99062  &297.34\\
KaSPar fast, $\alpha=\infty$   & 2\,697 & 2\,780 & 23.21 & -& - & -& - &- &-\\\hline
KaPPa strong  & 2\,807 & 2\,890 & 2.10 & 13\,323& 13\,600 & 28.16& 117701 &123613 &78.00 \\
KaPPa fast    & 2\,819 & 2\,910 & 1.29 & 13\,442& 13\,727 & 16.67& 117927 &126914 &46.40
\\\hline
kMetis        & 3\,097 & 3\,348 & 0.07 & 15\,540& 16\,656 &  0.71& 117959 & 134803 & 1.42 \\
Scotch        & 2\,926 & 3\,065 & 0.48 & 14\,475& 15\,074 &  3.83& 168764 & 168764 & 17.69 
\end{tabular}\\
\vspace*{5mm}
\begin{tabular}{r|r|r|r|r|r|r}
\hline
\multicolumn{7}{c}{Large Instances}\\\hline
k & \multicolumn{3}{c|}{KaSPar strong} & \multicolumn{3}{c}{KaPPa strong}\\
\hline
& best & avg. & t[s] & best & avg. & t[s] \\
\hline
2& 2\,842& 2\,873& 36.89& 2\,977& 3\,054& 15.03\\
4& 5\,642& 5\,707& 60.66& 6\,190& 6\,384& 30.31\\
8& 10\,464& 10\,580& 75.92& 11\,375& 11\,652& 37.86\\
16& 17\,345& 17\,567& 102.52& 18678& 19\,061& 39.13\\
32& 27\,416& 27\,707& 137.08& 29\,156& 29\,562& 31.35\\
64& 41\,284& 41\,570& 170.54& 43\,237& 43\,644& 22.36
\end{tabular}
\end{center}
\vspace{-7mm}
\end{table}

\paragraph*{The Walshaw Benchmark} \cite{SWC04plus} considers 34 graphs using $k\in \set{2, 4, 8, 16, 32, 64}$ and balance parameter $\epsilon\in\set{0,0.01,0.03,0.05}$ giving a total of 816 table entries.
Only cut sizes count -- running time is not reported.
We tried all combinations except the case $\epsilon=0$ which \algname\ cannot handle yet. We ran \algname\ strong with a time limit of one hour and report
the best result obtained in the appendix. \algname\ improved 155 values in the benchmark table: 42 for 1\%, 49 for 3\% and 64 for 5\% allowed imbalance. Moreover, it reproduced equally sized cuts in 83 additional cases.
If we count only results for graphs having over $44k$ nodes and $\epsilon>0$, \algname\ improved 131 and reproduced 27 cuts, thus summing up to 63\% of large graph table slots. We should note, that $51$ of the new improvements are over partitioners different from KaPPa.
Most of the improvements lie in the lower triangular part of the table, meaning  that \algname\ is particularly good for either large graphs, or smaller graphs with small $k$. On the other hand, for small graphs, large $k$, and $\epsilon = 1\%$ \algname\ was often not able to obtain a feasible solution. A primary reason for this seems to be that initial partitioning yields highly infeasible solutions that \algname\ is not able to to improve considerably during refinement. This is not astonishing, since Scotch targets $\epsilon=3\%$ and does not even guarantee that.\frage{discussion shortened}


\section{Conclusion}
\label{s:conclusion}

$n$-GP is a graph partitioning approach that scales to large inputs and currently computes the best known partitions for many large graphs, at least when a certain imbalance is allowed. It is in some sense simpler than previous methods since no matching algorithm is needed. Although our current implementation of \algname\ is a considerable constant factor slower than the fastest available MGP partitioners, we see potential for further tuning. In particular, thanks to our adaptive stopping rule, \algname\ needs to do very little local search, in particular for large graphs and small $k$. Thus it suffices to tune the relatively simple contraction routine to obtain significant improvements.  On the other hand, the adaptive stopping rule might also turn out to be useful for matching based MGP algorithms.

A lot of opportunities remain to further improve \algname. In particular,
we did not yet attempt to handle the case $\epsilon=0$ since this may require
different local search strategies. We also want to try other initial partitioning 
algorithms and ways to integrate $n$-GP into other metaheuristics like evolutionary search.

We expect that $n$-GP could be generalized for other objective functions, for
hypergraphs, and for graph clustering. More generally, the success of $n$-GP
also suggests to look for more applications of the $n$-level paradigm.

An apparent drawback of $n$-GP is that it looks inherently sequential. However, we
could probably obtain a good parallel algorithm by contracting \emph{small} sets
of highly rated, independent edges in parallel.  Indeed, in the light of our
results for \algname\ the complications coming from the need to find maximal
matchings of heavy edges seem unnecessary, i.e., a parallelization of $n$-GP
might be fast and simple. 

\paragraph*{Acknowledgements.} We would like to thank Christian Schulz for
supplying data for KaPPa, Scotch and Metis.

\bibliographystyle{abbrv}
\bibliography{diss,references}
\clearpage
\begin{appendix}
\section{Derivation of Stopping Rules}\label{app:stop}
Consider a situation where $p$ steps of local search have been performed
with average value $\mu$ and variance $\Variance$.
Then in the next $s$ steps, we can
expect a deviation from the expectation $(p+s)\mu$ by something of the order
$\sqrt{s\Variance}$. The expression $(p+s)\mu+\sqrt{s\Variance}$ is maximized
for $s^*\Is \frac{\Variance}{4\mu^2}$. Now the idea is to stop when for some tuning
parameter $x$, $(p+s^*)\mu+x\sqrt{s^*\Variance}>0$, i.e., it is reasonably
likely that a random walk modelling our local search can still give an
improvement. This translates to the condition
$p>\frac{\Variance}{\mu^2}(\frac{x}{2}-\frac{1}{4})$ or simply
$p\mu^2\gg\Variance$.
\section{Additional Figures and Tables}
\begin{figure}[h]
\begin{center}
\includegraphics[scale=0.8]{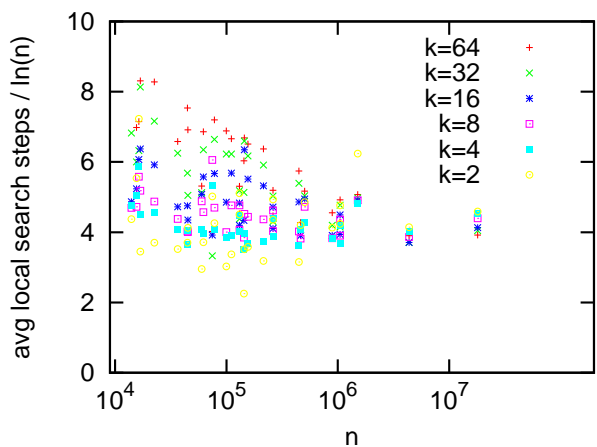}
\end{center}
\vspace*{-7mm}
\caption{Average length of local searches.}
\label{fig:localSearchLength}
\end{figure}
\begin{figure}[h]
\begin{center}
\includegraphics[scale=0.8]{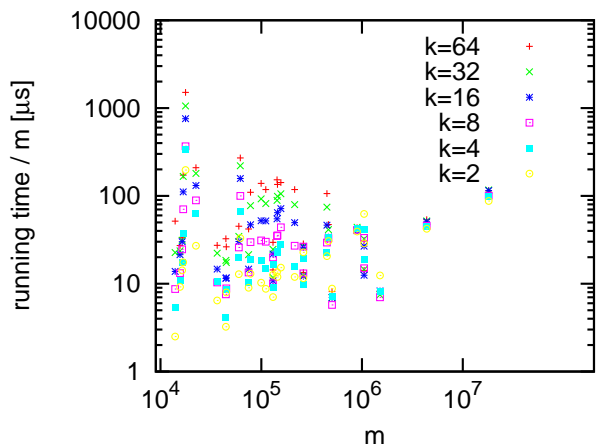}
\end{center}
\vspace*{-7mm}
\caption{Running Time.}
\label{fig:runTime}
\end{figure}
\begin{landscape}
\begin{table}[p]

\tiny

\hspace*{-2.5cm}\begin{tabular}{|l|r|r|r|r|r|r|r|r|r|r|r|r|r|r|r|r|r|r|r|r|r|}
 \hline

Graph  & k & \multicolumn{3}{|c|}{KaSPar fast} & \multicolumn{3}{|c|}{KaPPa strong} &
 \multicolumn{3}{|c|}{KaPPa fast} & \multicolumn{3}{|c|}{KaPPa minimal} & \multicolumn{3}{|c|}{scotch} & \multicolumn{3}{|c|} {metis}\\
\hline
& & best & avg & time & best & avg & time & best & avg & time & best & avg & time & best & avg & time & best & avg & time\\
\hline
coAuthorsCiteseer& 2& 17855& 18003& 25.77& 21775& 26462& 12.93& 29894& 30997& 11.98& 32678& 35492& 6.75& 34065& 34065& 5.34& 21587& 22674& 0.30\\
\hline
coAuthorsCiteseer& 4& 34180& 35315& 51.75& 43778& 46540& 28.43& 44837& 47156& 17.03& 50845& 55514& 5.78& 52277& 52277& 7.51& 39649& 41560& 0.33\\
\hline
coAuthorsCiteseer& 8& 49574& 50054& 85.49& 56574& 57647& 46.35& 53838& 55686& 25.77& 61397& 62752& 5.10& 69988& 69988& 9.61& 56289& 56996& 0.36\\
\hline
coAuthorsCiteseer& 16& 59574& 59915& 124.51& 66173& 66648& 55.26& 62126& 63085& 31.04& 65681& 67007& 5.71& 83457& 83457& 11.41& 68295& 68744& 0.39\\
\hline
coAuthorsCiteseer& 32& 67953& 68752& 169.78& 72331& 72736& 64.53& 71603& 72062& 30.34& 74119& 74760& 5.49& 90807& 90807& 12.92& 77399& 78254& 0.41\\
\hline
coAuthorsCiteseer& 64& 76210& 77326& 193.85& 77603& 78756& 64.45& 79411& 79872& 26.47& 81773& 82244& 5.82& 100737& 100737& 14.20& 84538& 85426& 0.44\\
\hline
citationCiteseer& 2& 32181& 32247& 49.44& 35122& 36696& 24.37& 34858& 48466& 15.31& 47641& 61055& 11.51& 37175& 37175& 5.87& 33684& 34344& 0.67\\
\hline
citationCiteseer& 4& 67194& 68371& 135.82& 76897& 79782& 47.87& 76994& 101369& 27.40& 120656& 133916& 12.54& 79543& 79543& 11.02& 73536& 77524& 0.76\\
\hline
citationCiteseer& 8& 103743& 105663& 297.70& 119852& 126129& 85.92& 118505& 133337& 47.09& 188731& 196204& 13.41& 124441& 124441& 15.37& 108655& 116082& 0.83\\
\hline
citationCiteseer& 16& 148932& 151256& 507.69& 156984& 164984& 114.26& 156132& 160555& 57.02& 218710& 224851& 14.18& 163941& 163941& 18.83& 153846& 157000& 0.91\\
\hline
citationCiteseer& 32& 198757& 203173& 841.73& 205922& 207923& 147.02& 198771& 207089& 111.12& 248894& 259090& 45.56& 210957& 210957& 21.88& 197146& 200650& 0.98\\
\hline
citationCiteseer& 64& 255722& 258037& 1213.93& 247462& 248994& 148.58& 240660& 241980& 115.18& 270692& 279247& 39.04& 265971& 265971& 25.08& 242010& 244427& 1.10\\
\hline
coAuthorsDBLP& 2& 45292& 45650& 82.42& 54803& 56140& 23.13& 55263& 61619& 16.96& 63305& 64872& 11.78& 63368& 63368& 8.23& 48952& 50341& 0.53\\
\hline
coAuthorsDBLP& 4& 80408& 81575& 144.65& 94651& 97597& 54.59& 97007& 98865& 32.09& 123373& 126675& 9.52& 109856& 109856& 11.73& 88513& 88734& 0.61\\
\hline
coAuthorsDBLP& 8& 109940& 113575& 263.08& 126261& 128129& 82.57& 116839& 118190& 46.32& 144839& 147038& 8.62& 142749& 142749& 14.49& 115201& 117074& 0.69\\
\hline
coAuthorsDBLP& 16& 132067& 135259& 440.42& 144229& 145229& 98.64& 137946& 138968& 46.24& 152803& 154368& 7.51& 169706& 169706& 16.97& 138399& 140149& 0.75\\
\hline
coAuthorsDBLP& 32& 152146& 154501& 787.16& 157754& 159086& 113.74& 151883& 153606& 62.94& 160331& 161779& 14.80& 189201& 189201& 18.95& 160842& 161565& 0.82\\
\hline
coAuthorsDBLP& 64& 168939& 169122& 1099.75& 169681& 170403& 123.69& 169283& 169671& 46.08& 174708& 175679& 10.11& 207486& 207486& 20.70& 175660& 177172& 0.88\\
\hline
cnr2000& 2& 210& 236& 49.54& 2597& 3789& 25.90& 2430& 4835& 23.21& 2422& 5053& 18.48& 20537& 20537& 3.70& 1773& 2451& 7.44\\
\hline
cnr2000& 4& 1569& 1973& 64.02& 6089& 6971& 46.71& 6284& 6939& 27.50& 6175& 7138& 13.34& 26809& 26809& 6.20& 4301& 6326& 8.05\\
\hline
cnr2000& 8& 4096& 4974& 75.55& 7914& 8510& 55.07& 7378& 7911& 33.20& 7684& 8308& 12.67& 31373& 31373& 8.18& 7899& 16951& 8.85\\
\hline
cnr2000& 16& 6943& 14824& 92.57& 8784& 10382& 61.45& 9399& 9567& 32.72& 9805& 10003& 12.45& 34967& 34967& 10.84& 12601& 81752& 9.47\\
\hline
cnr2000& 32& 384058& 400272& 188.69& 360661& 363687& 88.28& 368182& 372503& 48.97& 374786& 375787& 14.89& 432813& 432813& 12.59& 368062& 409130& 9.78\\
\hline
cnr2000& 64& 713772& 723710& 483.68& 694270& 700504& 103.42& 706366& 712434& 52.27& 722754& 727917& 15.49& 727685& 727685& 14.06& 723221& 737874& 10.31\\
\hline
coPapersDBLP& 2& 462530& 466947& 372.39& 512389& 527205& 80.25& 490054& 552438& 66.76& 528953& 585647& 60.58& 622378& 622378& 42.06& 599794& 634286& 2.33\\
\hline
coPapersDBLP& 4& 822518& 838005& 705.59& 937267& 952505& 122.23& 1021741& 1034181& 88.66& 1409276& 1428094& 42.62& 1188052& 1188052& 76.19& 1073007& 1091355& 2.58\\
\hline
coPapersDBLP& 8& 1188694& 1213398& 1794.77& 1257622& 1293223& 201.41& 1296044& 1315313& 131.60& 1690906& 1751688& 32.16& 1685436& 1685436& 98.29& 1442079& 1495943& 2.81\\
\hline
coPapersDBLP& 16& 1534078& 1544591& 3993.84& 1540054& 1571957& 318.56& 1593642& 1614871& 165.98& 1816467& 1852634& 28.73& 2028374& 2028374& 131.89& 1864836& 1886340& 3.01\\
\hline
coPapersDBLP& 32& 1789129& 1798109& 6550.18& 1828015& 1850535& 411.34& 1790694& 1861113& 276.51& 1926975& 2009450& 37.35& 2380424& 2380424& 156.07& 2087868& 2122569& 3.17\\
\hline
coPapersDBLP& 64& 2039271& 2054249& 10897.41& 2164396& 2177596& 423.03& 2051766& 2061702& 244.46& 2132793& 2139541& 31.87& 2697328& 2697328& 148.72& 2341150& 2347850& 3.39\\
\hline
\end{tabular}
\caption{All results for social network instances}
\end{table}	

\end{landscape}

\begin{landscape}
\pagestyle{empty}
\begin{table}[p]
 \tiny

\vspace*{-2.0cm}\hspace*{-3.25cm}\begin{tabular}{|l|r|r|r|r|r|r|r|r|r|r|r|r|r|r|r|r|r|r|r|r|r|r|r|r|}
\hline
Graph  & k & \multicolumn{3}{|c|}{KaSPar strong} & \multicolumn{3}{|c|}{KaSPar fast} & \multicolumn{3}{|c|}{KaPPa strong} & \multicolumn{3}{|c|}{KaPPa fast} & \multicolumn{3}{|c|}{KaPPa minimal} & \multicolumn{3}{|c|}{scotch} & \multicolumn{3}{|c|} {metis}\\
\hline
& &best & avg & time & best & avg & time & best & avg & time & best & avg & time & best & avg & time & best & avg & time & best & avg & time \\

\hline
fe\_tooth& 2& 3844& 3987& 5.86& 3840& 3981& 1.16& 3951& 4336& 3.75& 3854& 4353& 2.44& 4109& 6490& 1.59& 4259& 4259& 0.38& 4372& 4529& 0.08\\
\hline
fe\_tooth& 4& 6937& 6999& 8.54& 7034& 7146& 1.49& 7012& 7189& 5.22& 7126& 7757& 2.97& 7780& 9157& 0.96& 8304& 8304& 0.72& 7805& 8280& 0.08\\
\hline
fe\_tooth& 8& 11482& 11564& 13.43& 11574& 12007& 1.96& 12272& 12721& 6.83& 12215& 12678& 4.06& 13243& 13671& 0.75& 12999& 12999& 1.09& 13334& 13768& 0.08\\
\hline
fe\_tooth& 16& 17744& 17966& 21.24& 17968& 18288& 2.79& 18302& 18570& 7.18& 18198& 18524& 3.55& 19559& 19813& 0.65& 20816& 20816& 1.59& 20035& 20386& 0.09\\
\hline
fe\_tooth& 32& 25888& 26248& 35.12& 26249& 26592& 4.03& 26397& 26617& 5.28& 26404& 26677& 2.92& 28070& 28391& 0.53& 28430& 28430& 2.13& 28547& 29052& 0.10\\
\hline
fe\_tooth& 64& 36259& 36469& 49.65& 36741& 40385& 5.80& 36862& 37002& 4.71& 36795& 36992& 2.57& 38423& 39095& 0.62& 38401& 38401& 2.69& 39233& 39381& 0.12\\
\hline
598a& 2& 2371& 2384& 6.50& 2378& 2389& 1.84& 2387& 2393& 5.64& 2391& 2401& 3.79& 2456& 2485& 2.99& 2417& 2417& 0.39& 2444& 2513& 0.14\\
\hline
598a& 4& 7897& 7921& 11.15& 7935& 7977& 2.42& 8235& 8291& 10.24& 8291& 8385& 5.94& 9224& 9862& 2.62& 8246& 8246& 0.95& 8466& 8729& 0.15\\
\hline
598a& 8& 15929& 15984& 22.31& 15992& 16125& 3.48& 16502& 16641& 12.21& 16461& 16598& 7.07& 17351& 17899& 2.98& 17490& 17490& 1.63& 17170& 17533& 0.16\\
\hline
598a& 16& 26046& 26270& 38.39& 26102& 26672& 5.05& 26467& 26825& 17.74& 26670& 26887& 12.51& 27983& 28596& 6.76& 29804& 29804& 2.37& 27857& 28854& 0.17\\
\hline
598a& 32& 39625& 40019& 60.60& 40563& 40986& 7.25& 40946& 41190& 18.16& 40928& 41186& 11.91& 43111& 43741& 7.74& 44756& 44756& 3.21& 43256& 44213& 0.19\\
\hline
598a& 64& 58362& 58945& 87.52& 58326& 59199& 10.72& 59148& 59387& 14.15& 59026& 59233& 9.64& 61396& 61924& 6.21& 64561& 64561& 4.11& 61888& 62703& 0.22\\
\hline
fe\_ocean& 2& 317& 317& 5.55& 317& 322& 1.66& 314& 317& 3.21& 314& 318& 2.11& 343& 355& 1.71& 402& 402& 0.18& 540& 579& 0.11\\
\hline
fe\_ocean& 4& 1801& 1810& 9.40& 1817& 1837& 1.95& 1756& 1822& 6.30& 1754& 1822& 3.03& 1990& 2051& 1.10& 2000& 2000& 0.44& 2102& 2140& 0.11\\
\hline
fe\_ocean& 8& 4044& 4097& 14.33& 4084& 4195& 2.51& 4104& 4252& 6.33& 4143& 4330& 2.93& 4689& 4987& 0.73& 4956& 4956& 0.81& 5256& 5472& 0.12\\
\hline
fe\_ocean& 16& 7992& 8145& 22.41& 8120& 8359& 3.39& 8188& 8350& 5.62& 8294& 8469& 3.04& 9457& 9553& 0.70& 9351& 9351& 1.27& 10115& 10377& 0.13\\
\hline
fe\_ocean& 32& 13320& 13518& 36.53& 13526& 13806& 5.00& 13593& 13815& 4.34& 13618& 14042& 2.15& 15465& 15657& 0.47& 15089& 15089& 1.83& 16565& 16877& 0.15\\
\hline
fe\_ocean& 64& 21326& 21739& 62.46& 22059& 22209& 7.78& 21636& 21859& 3.68& 21809& 21973& 2.02& 24147& 24275& 0.51& 23246& 23246& 2.49& 24198& 24531& 0.17\\
\hline
144& 2& 6455& 6507& 12.81& 6461& 6491& 3.04& 6559& 6623& 7.45& 6563& 6638& 5.23& 6747& 6799& 3.64& 6695& 6695& 0.66& 6804& 6972& 0.20\\
\hline
144& 4& 15312& 15471& 24.73& 15717& 15774& 4.10& 16870& 16963& 13.33& 16998& 17122& 7.00& 17364& 18101& 2.97& 16899& 16899& 1.44& 17144& 17487& 0.21\\
\hline
144& 8& 25130& 25409& 38.13& 25557& 26039& 5.54& 26300& 26457& 20.11& 26435& 26614& 10.49& 27206& 27829& 2.93& 28172& 28172& 2.24& 28006& 28194& 0.22\\
\hline
144& 16& 37872& 38404& 69.35& 38830& 39161& 8.30& 39010& 39319& 26.04& 39266& 39492& 17.53& 40264& 41977& 6.63& 43712& 43712& 3.12& 42861& 43041& 0.24\\
\hline
144& 32& 57082& 57492& 106.40& 57353& 57860& 11.73& 58331& 58631& 24.60& 58175& 58652& 16.03& 61774& 62171& 8.79& 63224& 63224& 4.14& 61716& 62481& 0.26\\
\hline
144& 64& 80313& 80770& 144.77& 80609& 81293& 16.05& 82286& 82452& 19.11& 82029& 82493& 12.05& 86067& 86950& 8.16& 88246& 88246& 5.25& 86534& 87208& 0.30\\
\hline
wave& 2& 8661& 8720& 16.19& 8650& 8690& 3.25& 8832& 9132& 8.24& 8809& 9108& 4.72& 8966& 9324& 2.84& 9337& 9337& 0.83& 9169& 9345& 0.19\\
\hline
wave& 4& 16806& 16920& 29.56& 16871& 16978& 4.39& 17008& 17250& 14.51& 17263& 17503& 6.84& 18041& 21189& 1.81& 19995& 19995& 1.72& 19929& 21906& 0.20\\
\hline
wave& 8& 28681& 28817& 46.61& 28865& 29200& 6.01& 30690& 31419& 20.63& 30628& 31371& 9.79& 32617& 33937& 1.50& 33357& 33357& 2.61& 33223& 33639& 0.21\\
\hline
wave& 16& 42918& 43208& 75.97& 43267& 43770& 8.31& 44831& 45048& 20.54& 44936& 45202& 10.73& 46293& 47270& 1.47& 48903& 48903& 3.53& 48404& 49000& 0.22\\
\hline
wave& 32& 63025& 63159& 112.19& 62764& 63266& 11.88& 63981& 64390& 14.94& 64004& 64532& 8.19& 68085& 68620& 1.02& 70581& 70581& 4.68& 68062& 68604& 0.25\\
\hline
wave& 64& 87243& 87554& 150.37& 87403& 87889& 16.88& 88376& 88964& 12.51& 88924& 89297& 6.09& 92366& 93424& 1.03& 96759& 96759& 5.90& 92148& 94083& 0.29\\
\hline
m14b& 2& 3828& 3846& 20.03& 3845& 3870& 4.49& 3862& 3954& 11.16& 3900& 3945& 7.80& 3951& 4208& 5.63& 3872& 3872& 0.70& 4036& 4155& 0.31\\
\hline
m14b& 4& 13015& 13079& 26.51& 13111& 13160& 5.42& 13543& 13810& 18.77& 14104& 14211& 10.21& 14990& 15094& 4.42& 13484& 13484& 1.71& 13932& 14560& 0.33\\
\hline
m14b& 8& 25573& 25756& 45.33& 25839& 26086& 7.27& 27330& 27393& 24.97& 27411& 27450& 11.76& 28241& 28517& 3.78& 27839& 27839& 2.86& 28138& 28507& 0.34\\
\hline
m14b& 16& 42212& 42458& 83.25& 42727& 43365& 10.07& 45352& 45762& 28.11& 45931& 46108& 19.27& 48769& 49397& 6.41& 50778& 50778& 4.25& 48314& 49269& 0.36\\
\hline
m14b& 32& 66314& 66991& 133.88& 66942& 68017& 14.52& 68107& 69075& 29.94& 68715& 69223& 17.99& 72484& 73598& 8.12& 75453& 75453& 5.72& 72746& 74135& 0.40\\
\hline
m14b& 64& 99207& 100014& 198.23& 99964& 100666& 20.91& 101053& 101455& 25.26& 101410& 101861& 17.46& 106361& 107173& 10.24& 109404& 109404& 7.38& 107384& 108141& 0.44\\
\hline
auto& 2& 9740& 9768& 68.39& 9744& 9776& 10.99& 9910& 10045& 30.09& 9863& 10856& 18.86& 10313& 11813& 12.12& 10666& 10666& 1.61& 10781& 12147& 0.83\\
\hline
auto& 4& 25988& 26062& 75.60& 26072& 26116& 13.35& 28218& 29481& 64.01& 29690& 29995& 33.11& 32473& 33371& 8.93& 29046& 29046& 3.52& 27469& 30318& 0.86\\
\hline
auto& 8& 45099& 45232& 97.60& 45416& 45806& 15.98& 46272& 46652& 85.89& 47163& 48229& 46.36& 49447& 53617& 8.42& 49999& 49999& 5.42& 49691& 52422& 0.87\\
\hline
auto& 16& 76287& 76715& 153.46& 77376& 77801& 20.81& 78713& 79769& 87.41& 79711& 80683& 58.20& 84236& 86001& 12.25& 84462& 84462& 7.84& 85562& 89139& 0.91\\
\hline
auto& 32& 121269& 121862& 246.50& 122406& 123052& 28.12& 124606& 125500& 71.77& 124920& 125876& 46.44& 131545& 133723& 20.23& 133403& 133403& 10.58& 133026& 134086& 0.99\\
\hline
auto& 64& 174612& 174914& 352.09& 174712& 176214& 38.76& 177038& 177595& 62.64& 177461& 178119& 44.14& 185836& 187424& 25.39& 193170& 193170& 13.68& 188555& 189699& 1.08\\
\hline
delaunay\_n20& 2& 1711& 1731& 196.33& 1726& 1753& 12.88& 1858& 1882& 35.43& 1879& 1898& 18.66& 1911& 1937& 13.87& 1874& 1874& 1.18& 2054& 2194& 1.11\\
\hline
delaunay\_n20& 4& 3418& 3439& 130.67& 3460& 3480& 13.21& 3674& 3780& 64.08& 3784& 3826& 24.34& 3857& 3900& 8.97& 3723& 3723& 2.35& 4046& 4094& 1.15\\
\hline
delaunay\_n20& 8& 6278& 6317& 104.37& 6364& 6387& 13.71& 6670& 6854& 70.07& 6688& 6872& 41.92& 7161& 7303& 6.15& 7180& 7180& 3.58& 7705& 8029& 1.13\\
\hline
delaunay\_n20& 16& 10183& 10218& 84.33& 10230& 10327& 13.80& 10816& 11008& 67.92& 10882& 11061& 48.05& 11307& 11533& 6.31& 11266& 11266& 4.77& 11854& 12440& 1.14\\
\hline
delaunay\_n20& 32& 15905& 16026& 101.69& 16211& 16236& 14.90& 16813& 17086& 42.67& 16814& 17150& 24.44& 17993& 18179& 3.33& 17784& 17784& 6.04& 18816& 19304& 1.18\\
\hline
delaunay\_n20& 64& 23935& 23962& 97.09& 24193& 24263& 16.40& 24799& 25179& 22.04& 24946& 25129& 12.83& 26314& 27001& 1.79& 26163& 26163& 7.34& 28318& 28543& 1.21\\
\hline
rgg\_n\_2\_20\_s0& 2& 2162& 2201& 198.61& 2146& 2217& 16.75& 2377& 2498& 33.24& 2378& 2497& 24.66& 2400& 2530& 20.71& 2832& 2832& 1.41& 3023& 3326& 1.57\\
\hline
rgg\_n\_2\_20\_s0& 4& 4323& 4389& 130.00& 4382& 4448& 17.18& 4867& 5058& 38.50& 4870& 4973& 21.06& 5114& 5200& 11.11& 5737& 5737& 2.82& 5786& 6174& 1.56\\
\hline
rgg\_n\_2\_20\_s0& 8& 7745& 7915& 103.66& 8031& 8174& 17.81& 8995& 9391& 46.06& 9248& 9493& 25.50& 9426& 9632& 7.83& 11251& 11251& 4.48& 11365& 11771& 1.54\\
\hline
rgg\_n\_2\_20\_s0& 16& 12596& 12792& 86.19& 12981& 13148& 17.93& 14953& 15199& 35.86& 15013& 15339& 24.61& 15039& 15442& 7.20& 17157& 17157& 6.13& 17498& 18125& 1.53\\
\hline
rgg\_n\_2\_20\_s0& 32& 20403& 20478& 100.03& 20805& 20958& 18.99& 23430& 23917& 26.04& 23383& 24222& 16.93& 23842& 24164& 3.94& 28078& 28078& 7.96& 27765& 28495& 1.58\\
\hline
rgg\_n\_2\_20\_s0& 64& 30860& 31066& 97.83& 31203& 31584& 20.50& 34778& 35354& 11.62& 35086& 35539& 9.95& 35252& 35629& 2.09& 38815& 38815& 9.83& 41066& 42465& 1.58\\
\hline
af\_shell10& 2& 26225& 26225& 317.11& 26225& 26225& 37.00& 26225& 26225& 78.65& 26225& 26225& 65.31& 26525& 26640& 59.76& 26825& 26825& 3.64& 27625& 28955& 2.99\\
\hline
af\_shell10& 4& 55075& 55345& 210.61& 55875& 56375& 36.59& 54950& 55265& 91.96& 54950& 55500& 51.52& 58366& 58627& 22.11& 58500& 58500& 7.60& 61100& 64705& 3.04\\
\hline
af\_shell10& 8& 97709& 100233& 179.51& 100325& 102667& 38.47& 101425& 102335& 136.99& 102125& 103180& 61.16& 110369& 111081& 16.03& 105375& 105375& 11.97& 117650& 120120& 3.04\\
\hline
af\_shell10& 16& 163125& 165770& 212.12& 163600& 165360& 40.47& 165025& 166427& 106.63& 165625& 166480& 69.97& 174677& 175918& 17.00& 171725& 171725& 16.45& 184350& 188765& 3.06\\
\hline
af\_shell10& 32& 248268& 252939& 191.53& 252555& 256262& 43.14& 253525& 255535& 80.85& 252487& 255746& 52.00& 270249& 275149& 9.25& 269375& 269375& 21.66& 289400& 291590& 3.13\\
\hline
af\_shell10& 64& 372823& 376512& 207.76& 378031& 382191& 49.38& 379125& 382923& 43.01& 380225& 384140& 29.43& 400378& 404085& 4.82& 402275& 402275& 27.33& 421285& 427047& 3.18\\
\hline
deu& 2& 167& 172& 231.47& 175& 179& 58.31& 214& 221& 68.20& 230& 240& 47.55& 233& 243& 38.11& 295& 295& 3.19& 268& 286& 5.38\\
\hline
deu& 4& 419& 426& 244.12& 427& 447& 58.84& 533& 542& 76.87& 531& 545& 49.37& 544& 580& 25.65& 726& 726& 6.46& 699& 761& 5.35\\
\hline
deu& 8& 762& 773& 250.50& 781& 792& 59.20& 922& 962& 99.76& 935& 973& 45.05& 974& 1007& 19.57& 1235& 1235& 9.84& 1174& 1330& 5.24\\
\hline
deu& 16& 1308& 1333& 278.31& 1332& 1387& 61.82& 1550& 1616& 105.96& 1556& 1618& 78.82& 1593& 1656& 21.79& 2066& 2066& 13.11& 2041& 2161& 5.19\\
\hline
deu& 32& 2182& 2217& 283.79& 2251& 2295& 62.50& 2548& 2615& 73.17& 2535& 2641& 41.93& 2626& 2711& 11.50& 3250& 3250& 16.28& 3319& 3445& 5.28\\
\hline
deu& 64& 3610& 3631& 293.53& 3679& 3737& 64.38& 4021& 4093& 49.55& 4078& 4146& 31.63& 4193& 4317& 5.97& 4978& 4978& 19.41& 5147& 5385& 5.31\\
\hline
eur& 2& 133& 138& 1946.34& 162& 211& 792.68& & & & & & & & & & 469& 469& 12.45& & & \\
\hline
eur& 4& 355& 375& 2168.10& 416& 431& 794.41& 543& 619& 441.11& 580& 646& 223.96& 657& 697& 113.35& 952& 952& 25.37& 846& 1626& 29.40\\
\hline
eur& 8& 774& 786& 2232.31& 823& 834& 809.21& 986& 1034& 418.29& 1013& 1034& 207.41& 1060& 1119& 80.92& 1667& 1667& 38.67& 1675& 3227& 29.04\\
\hline
eur& 16& 1401& 1440& 2553.40& 1575& 1597& 930.59& 1760& 1900& 497.93& 1907& 1935& 295.81& 1931& 2048& 94.56& 2922& 2922& 51.50& 3519& 9395& 30.58\\
\hline
eur& 32& 2595& 2643& 2598.84& 2681& 2761& 958.24& 3186& 3291& 417.52& 3231& 3314& 306.52& 3202& 3386& 55.63& 4336& 4336& 65.16& 7424& 9442& 30.81\\
\hline
eur& 64& 4502& 4526& 2533.56& 4622& 4675& 868.75& 5290& 5393& 308.17& 5448& 5538& 183.98& 5569& 5770& 29.64& 6772& 6772& 77.14& 11313& 12738& 30.30\\
\hline
\end{tabular}
\caption{All results for large instances.\label{tab:all}}
\end{table}
\end{landscape}

\begin{table}[h]
\begin{center}\scriptsize
\hspace*{-0.8cm}\begin{tabular}{|l|r|r|r|r|r|r|r|r|r|r|r|r|}\hline

Graph  & \multicolumn{2}{|c|}{2} & \multicolumn{2}{|c|}{4} & \multicolumn{2}{|c|}{8} & \multicolumn{2}{|c|}{16} & \multicolumn{2}{|c|}{32} & \multicolumn{2}{|c|}{64}\\

\hline
add20& 641& 594& 1212& 1177& 1814& 1704& 2427& 2121& & 2687& & 3236\\
\hline
data& 190& 188& 405& 383& 699& 660& & 1162& & 1865& & 2885\\
\hline
3elt& 90& 89& 201& 199& 361& 342& 654& 569& & 969& & 1564\\
\hline
uk& \textbf{19}& 19& \textbf{41}& 42& 92& 84& 179& 152& & 258& & 438\\
\hline
add32& \textbf{10}& 10& \textbf{33}& 33& \textbf{66}& 66& \textbf{117}& 117& \textbf{212}& 212& & 493\\
\hline
bcsstk33& 10105& 10097& 21756& 21508& 34377& 34178& 56687& 54860& & 78132& & 108505\\
\hline
whitaker3& \textbf{126}& 126& 382& 380& 670& 656& 1163& 1093& & 1717& & 2567\\
\hline
crack& 184& 183& 370& 362& 696& 678& 1183& 1092& & 1707& & 2566\\
\hline
wing\_nodal& 1703& 1696& 3609& 3572& 5574& 5443& 8624& 8422& & 11980& & 16134\\
\hline
fe\_4elt2& \textbf{130}& 130& \textbf{349}& 349& 622& 605& 1051& 1014& & 1657& & 2537\\
\hline
vibrobox& 11538& 10310& 19267& 19199& 25190& 24553& 35514& 32167& 46331& 41399& & 49521\\
\hline
bcsstk29& \textbf{2818}& 2818& \textbf{8035}& 8159& 14212& 13965& 23808& 21768& & 34886& & 57054\\
\hline
4elt& \textbf{138}& 138& 325& 321& 561& 534& 1009& 939& & 1559& & 2596\\
\hline
fe\_sphere& \textbf{386}& 386& 798& 768& 1236& 1152& 1914& 1730& & 2565& & 3663\\
\hline
cti& \textbf{318}& 318& 950& 944& 1815& 1802& 3056& 2906& 5044& 4223& & 5875\\
\hline
memplus& 5698& 5489& 10234& 9559& 12599& 11785& 14410& 13241& 16340& 14395& & 16857\\
\hline
cs4& 378& 367& 970& 940& 1520& 1467& 2285& 2206& 3521& 3090& & 4169\\
\hline
bcsstk30& 6347& 6335& \textbf{16617}& 16622& 34761& 34604& 72028& 71234& & 115770& & 173945\\
\hline
bcsstk31& 2723& 2701& \textbf{7351}& 7444& \textbf{13371}& 13417& 24791& 24277& 42745& 38086& & 60528\\
\hline
fe\_pwt& \textbf{340}& 340& \textbf{704}& 704& \textbf{1441}& 1442& 2835& 2806& & 5612& & 8454\\
\hline
bcsstk32& \textbf{4667}& 4667& \textbf{9247}& 9492& \textbf{20855}& 21490& \textbf{37372}& 37673& 72471& 61144& & 95199\\
\hline
fe\_body& \textbf{262}& 262& \textbf{599}& 636& \textbf{1079}& 1156& \textbf{1858}& 1931& & 3202& & 5282\\
\hline
t60k& 78& 75& 213& 211& 470& 465& 866& 849& 1493& 1391& & 2211\\
\hline
wing& 803& 787& 1683& 1666& 2616& 2589& 4147& 4131& 6271& 5902& & 8132\\
\hline
brack2& \textbf{708}& 708& \textbf{3027}& 3038& \textbf{7144}& 7269& \textbf{11969}& 11983& 18496& 17798& & 26557\\
\hline
finan512& \textbf{162}& 162& \textbf{324}& 324& \textbf{648}& 648& \textbf{1296}& 1296& \textbf{2592}& 2592& & 10560\\
\hline
fe\_tooth& \textbf{3819}& 3823& \textbf{6938}& 7103& \textbf{11650}& 11935& \textbf{18115}& 18283& 26604& 25977& & 35980\\
\hline
fe\_rotor& 2055& 2045& \textbf{7405}& 7480& \textbf{12959}& 13165& 21093& 20773& 33588& 32783& & 47461\\
\hline
598a& 2390& 2388& \textbf{7992}& 8154& \textbf{16179}& 16467& \textbf{26196}& 26427& \textbf{40513}& 40674& & 59098\\
\hline
fe\_ocean& 388& 387& \textbf{1856}& 1878& \textbf{4251}& 4299& \textbf{8276}& 8432& 13841& 13660& & 21548\\
\hline
144& 6489& 6479& \textbf{15196}& 15345& \textbf{25455}& 25818& \textbf{38940}& 39352& 58359& 58126& & 81145\\
\hline
wave& 8716& 8682& \textbf{16891}& 17475& \textbf{29207}& 30511& \textbf{43697}& 44611& \textbf{64198}& 64551& & 88863\\
\hline
m14b& 3828& 3826& \textbf{13034}& 13391& \textbf{25921}& 26666& \textbf{42513}& 43975& 67990& 67770& & 101551\\
\hline
auto& \textbf{10004}& 10042& \textbf{26941}& 27790& \textbf{45731}& 47650& \textbf{77618}& 79847& \textbf{123296}& 124991& 179309& 175975\\
\hline
\end{tabular}
\end{center}
\caption{Walshaw Benchmark with $\epsilon=1$}
\end{table}

\begin{table}[p]
\begin{center}\scriptsize
\hspace*{-0.8cm}\begin{tabular}{|l|r|r|r|r|r|r|r|r|r|r|r|r|}\hline

Graph  & \multicolumn{2}{|c|}{2} & \multicolumn{2}{|c|}{4} & \multicolumn{2}{|c|}{8} & \multicolumn{2}{|c|}{16} & \multicolumn{2}{|c|}{32} & \multicolumn{2}{|c|}{64}\\
\hline
add20& 636& 576& 1195& 1158& 1765& 1690& 2331& 2095& 2862& 2493& & 3152\\
\hline
data& 186& 185& 379& 378& 662& 650& 1163& 1133& 1972& 1802& & 2809\\
\hline
3elt& \textbf{87}& 87& 199& 198& 346& 336& 587& 565& 1035& 958& 1756& 1542\\
\hline
uk& \textbf{18}& 18& \textbf{40}& 40& 84& 81& 158& 148& 281& 251& 493& 414\\
\hline
add32& \textbf{10}& 10& \textbf{33}& 33& \textbf{66}& 66& \textbf{117}& 117& \textbf{212}& 212& 509& 493\\
\hline
bcsstk33& \textbf{10064}& 10064& 21083& 21035& 34150& 34078& 55372& 54510& 80548& 77672& 113269& 107012\\
\hline
whitaker3& \textbf{126}& 126& 381& 378& 662& 655& 1125& 1092& 1757& 1686& 2733& 2535\\
\hline
crack& \textbf{182}& 182& \textbf{360}& 360& 685& 676& 1132& 1082& 1765& 1679& 2739& 2553\\
\hline
wing\_nodal& 1681& 1680& 3572& 3561& 5424& 5401& 8476& 8316& 12282& 11938& 16891& 15971\\
\hline
fe\_4elt2& \textbf{130}& 130& 349& 343& 607& 598& 1022& 1007& 1686& 1633& 2658& 2527\\
\hline
vibrobox& 11538& 10310& 19239& 18778& 24691& 24171& 34226& 31516& 43532& 39592& 52242& 49123\\
\hline
bcsstk29& \textbf{2818}& 2818& \textbf{7983}& 8045& 14041& 13817& 22448& 21410& 35660& 34407& 58644& 55366\\
\hline
4elt& \textbf{137}& 137& \textbf{319}& 319& 533& 523& 942& 914& 1631& 1537& 2728& 2581\\
\hline
fe\_sphere& \textbf{384}& 384& 792& 764& 1193& 1152& 1816& 1706& 2715& 2477& 3965& 3547\\
\hline
cti& \textbf{318}& 318& 924& 917& 1724& 1716& 2900& 2778& 4396& 4132& 6330& 5763\\
\hline
memplus& 5626& 5355& 10145& 9418& 12521& 11628& 14168& 13130& 15850& 14264& 18364& 16724\\
\hline
cs4& 366& 361& 959& 936& 1490& 1467& 2215& 2126& 3152& 3048& 4479& 4169\\
\hline
bcsstk30& \textbf{6251}& 6251& \textbf{16497}& 16537& \textbf{34275}& 34513& 70851& 70278& 117500& 114005& 178977& 171727\\
\hline
bcsstk31& \textbf{2676}& 2676& 7183& 7181& \textbf{13090}& 13246& 24211& 23504& 39298& 37459& 60847& 58667\\
\hline
fe\_pwt& \textbf{340}& 340& \textbf{704}& 704& \textbf{1416}& 1419& 2787& 2784& 5649& 5606& 8557& 8346\\
\hline
bcsstk32& \textbf{4667}& 4667& \textbf{8778}& 8799& \textbf{20035}& 21023& \textbf{35788}& 36613& 61485& 59824& 96086& 92690\\
\hline
fe\_body& \textbf{262}& 262& \textbf{598}& 601& \textbf{1033}& 1054& \textbf{1767}& 1800& \textbf{2906}& 2947& \textbf{4982}& 5212\\
\hline
t60k& \textbf{71}& 71& 211& 207& 461& 454& 851& 822& 1423& 1391& 2264& 2198\\
\hline
wing& 789& 774& 1660& 1636& 2567& 2551& 4034& 4015& 6005& 5832& 8316& 8043\\
\hline
brack2& \textbf{684}& 684& 2853& 2839& \textbf{6980}& 6994& \textbf{11622}& 11741& \textbf{17491}& 17649& 26679& 26366\\
\hline
finan512& \textbf{162}& 162& \textbf{324}& 324& \textbf{648}& 648& \textbf{1296}& 1296& \textbf{2592}& 2592& 10635& 10560\\
\hline
fe\_tooth& 3794& 3792& \textbf{6862}& 6946& \textbf{11422}& 11662& \textbf{17655}& 17760& 25685& 25624& 35962& 35830\\
\hline
fe\_rotor& \textbf{1960}& 1963& \textbf{7182}& 7222& \textbf{12546}& 12852& \textbf{20356}& 20521& 32114& 31763& 47613& 47049\\
\hline
598a& 2369& 2367& \textbf{7873}& 7955& \textbf{15820}& 16031& \textbf{25927}& 25966& \textbf{39525}& 39829& \textbf{58101}& 58454\\
\hline
fe\_ocean& \textbf{311}& 311& 1710& 1698& 3976& 3974& 7919& 7838& 12942& 12746& 21217& 21033\\
\hline
144& 6456& 6438& \textbf{15122}& 15250& \textbf{25301}& 25491& \textbf{37899}& 38478& \textbf{56463}& 57354& \textbf{80621}& 80767\\
\hline
wave& 8640& 8616& \textbf{16822}& 16936& \textbf{28664}& 28839& \textbf{42620}& 43063& \textbf{62281}& 62743& \textbf{86663}& 87325\\
\hline
m14b& 3828& 3823& \textbf{12977}& 13136& \textbf{25550}& 26057& \textbf{42061}& 42783& \textbf{65879}& 67326& \textbf{98188}& 100286\\
\hline
auto& \textbf{9716}& 9782& \textbf{25979}& 26379& \textbf{45109}& 45525& \textbf{76016}& 77611& \textbf{120534}& 122902& \textbf{172357}& 174904\\
\hline
\end{tabular}
\end{center}
\caption{Walshaw Benchmark with $\epsilon=3$}
\end{table}

\begin{table}[p]
\begin{center}
\scriptsize
\hspace*{-0.8cm}\begin{tabular}{|l|r|r|r|r|r|r|r|r|r|r|r|r|}\hline

Graph  & \multicolumn{2}{|c|}{2} & \multicolumn{2}{|c|}{4} & \multicolumn{2}{|c|}{8} & \multicolumn{2}{|c|}{16} & \multicolumn{2}{|c|}{32} & \multicolumn{2}{|c|}{64}\\
\hline
add20& 610& 550& 1186& 1157& 1755& 1675& 2267& 2081& 2786& 2463& 3270& 3152\\
\hline
data& 183& 181& 369& 368& 640& 628& 1130& 1086& 1907& 1777& 3073& 2798\\
\hline
3elt& \textbf{87}& 87& 198& 197& 336& 330& 572& 560& 1009& 950& 1645& 1539\\
\hline
uk& \textbf{18}& 18& \textbf{39}& 40& 81& 78& 150& 139& 272& 246& 456& 410\\
\hline
add32& \textbf{10}& 10& \textbf{33}& 33& \textbf{63}& 65& \textbf{117}& 117& \textbf{212}& 212& \textbf{491}& 493\\
\hline
bcsstk33& \textbf{9914}& 9914& \textbf{20198}& 20584& 33971& 33938& 55273& 54323& 79159& 77163& 111659& 106886\\
\hline
whitaker3& \textbf{126}& 126& 380& 378& 658& 650& 1110& 1084& 1741& 1686& 2663& 2535\\
\hline
crack& \textbf{182}& 182& 361& 360& 673& 667& 1096& 1080& 1749& 1679& 2681& 2548\\
\hline
wing\_nodal& 1672& 1668& 3541& 3536& 5375& 5350& 8419& 8316& 12149& 11879& 16566& 15873\\
\hline
fe\_4elt2& \textbf{130}& 130& 340& 335& 596& 583& 1013& 991& 1665& 1633& 2608& 2516\\
\hline
vibrobox& 11538& 10310& 19021& 18778& 24203& 23930& 34298& 31235& 42890& 39592& 50994& 48200\\
\hline
bcsstk29& \textbf{2818}& 2818& \textbf{7936}& 7942& 13619& 13614& 21914& 20924& 34906& 33818& 57220& 54935\\
\hline
4elt& \textbf{137}& 137& 318& 315& 519& 516& 925& 902& 1574& 1532& 2673& 2565\\
\hline
fe\_sphere& \textbf{384}& 384& 784& 764& 1219& 1152& 1801& 1692& 2678& 2477& 3904& 3547\\
\hline
cti& \textbf{318}& 318& 900& 890& \textbf{1708}& 1716& 2830& 2725& 4227& 4037& 6127& 5684\\
\hline
memplus& 5516& 5267& 10011& 9299& 12458& 11555& 14047& 13078& 15749& 14170& 18213& 16454\\
\hline
cs4& 363& 356& 955& 936& 1483& 1467& 2184& 2126& 3115& 2995& 4394& 4116\\
\hline
bcsstk30& \textbf{6251}& 6251& \textbf{16186}& 16332& \textbf{34146}& 34350& \textbf{69520}& 70043& 114960& 113321& 175723& 170591\\
\hline
bcsstk31& \textbf{2676}& 2676& \textbf{7099}& 7152& \textbf{12941}& 13058& 23603& 23254& 38150& 37459& 60768& 57534\\
\hline
fe\_pwt& \textbf{340}& 340& \textbf{700}& 701& \textbf{1405}& 1409& \textbf{2772}& 2777& \textbf{5545}& 5546& 8410& 8310\\
\hline
bcsstk32& \textbf{4622}& 4644& \textbf{8454}& 8481& \textbf{19678}& 20099& \textbf{35208}& 35965& 60441& 59824& 94238& 91006\\
\hline
fe\_body& \textbf{262}& 262& \textbf{596}& 601& \textbf{1017}& 1054& \textbf{1723}& 1784& \textbf{2807}& 2887& \textbf{4834}& 4888\\
\hline
t60k& \textbf{65}& 65& 202& 196& 457& 454& 839& 818& 1398& 1376& 2229& 2168\\
\hline
wing& 784& 770& 1654& 1636& \textbf{2528}& 2551& \textbf{3998}& 4015& 5915& 5806& 8228& 7991\\
\hline
brack2& \textbf{660}& 660& 2745& 2739& \textbf{6671}& 6781& \textbf{11358}& 11558& \textbf{17256}& 17529& 26321& 26281\\
\hline
finan512& \textbf{162}& 162& \textbf{324}& 324& \textbf{648}& 648& \textbf{1296}& 1296& \textbf{2592}& 2592& 10583& 10560\\
\hline
fe\_tooth& 3780& 3773& \textbf{6825}& 6864& \textbf{11337}& 11662& \textbf{17404}& 17603& \textbf{25216}& 25624& \textbf{35466}& 35476\\
\hline
fe\_rotor& \textbf{1950}& 1955& 7052& 7045& \textbf{12380}& 12566& \textbf{20039}& 20132& \textbf{31450}& 31576& 46749& 46608\\
\hline
598a& 2338& 2336& \textbf{7763}& 7851& \textbf{15544}& 15721& \textbf{25585}& 25808& \textbf{39144}& 39369& \textbf{57412}& 58031\\
\hline
fe\_ocean& \textbf{311}& 311& 1705& 1697& 3946& 3941& \textbf{7618}& 7722& \textbf{12720}& 12746& 20886& 20667\\
\hline
144& 6373& 6362& \textbf{15036}& 15250& \textbf{25025}& 25259& \textbf{37433}& 38225& \textbf{56345}& 56926& \textbf{79296}& 80257\\
\hline
wave& 8598& 8563& \textbf{16662}& 16820& \textbf{28615}& 28700& \textbf{42482}& 42800& \textbf{61788}& 62520& \textbf{85658}& 86663\\
\hline
m14b& 3806& 3802& \textbf{12976}& 13136& \textbf{25292}& 25679& \textbf{41750}& 42608& \textbf{65231}& 66793& \textbf{98005}& 99063\\
\hline
auto& 9487& 9450& \textbf{25399}& 25883& \textbf{44520}& 45039& \textbf{75066}& 76488& \textbf{120001}& 122378& \textbf{171459}& 173968\\
\hline
\end{tabular}
\end{center}
\caption{Walshaw Benchmark with $\epsilon=5$}
\end{table}

\end{appendix}
\end{document}

%% file: makros2.tex
\newcommand{\Id}[1]{\ensuremath{\text{{\sf #1}}}}

\newcommand{\set}[1]{\left\{ #1\right\}}
\newcommand{\gilt}{:}

\newcommand{\setGilt}[2]{\left\{ #1\gilt #2\right\}}

\newcommand{\realrange}[2]{\left[#1, #2\right]}

\newcommand{\unitrange}[2]{\realrange{0}{1}}

\newcommand{\Oh}[1]{\mathcal{O}\!\left( #1\right)}

\newcommand{\llabel}[1]{\label{\labelprefix:#1}}
\newcommand{\labelprefix}{} 

\newcommand{\discussionsize}{\small}

\newcommand{\frage}[1]{{[\bf #1]}\marginpar{$\bigotimes$}}

\newenvironment{code}{\noindent\normalsize
\begin{tabbing}%
\hspace{2em}\=\hspace{2em}\=\hspace{2em}\=\hspace{2em}\=\hspace{2em}\=%
\hspace{2em}\=\hspace{2em}\=\hspace{2em}\=\hspace{2em}\=\hspace{2em}\=%
\kill}{\end{tabbing}}

\newcommand{\labelcommand}{}
\newcommand{\captiontext}{}
\newsavebox{\codeparam}
\newcounter{lineNumber}
\newenvironment{disscodepos}[3]{%
\renewcommand{\labelcommand}{#2}%
\renewcommand{\captiontext}{#3}%
\sbox{\codeparam}{\parbox{\textwidth}{#3}}%
\begin{figure}[#1]\begin{center}\begin{code}\setcounter{lineNumber}{1}}{%
\end{code}\end{center}\caption{\llabel{\labelcommand}\captiontext}\end{figure}}

{\end{disscodepos}}

\newcommand{\Function} {{\bf Function\ }}

\newcommand{\Is}{\mbox{\rm := }}

\newcommand{\If}       {{\bf if\ }}

\newcommand{\Then}     {{\bf then\ }}

\newcommand{\Return}   {{\bf return\ }}

\newdimen\endofsize\endofsize=0.5em